\documentclass[fleqn,twoside]{article}
\usepackage{espcrc2}

\usepackage{graphicx}
\usepackage[figuresright]{rotating}

\newcommand{\AmS}{{\protect\the\textfont2
  A\kern-.1667em\lower.5ex\hbox{M}\kern-.125emS}}

\title{The end of the galactic cosmic ray spectrum
}

\author{Todor Stanev\address{Bartol Research Institute, 
        Department of Physics and Astronomy, \\ 
        University of Delaware, Newark, DE~19716, U.S.A.}}
\begin{document}

\begin{abstract}
 We discuss the region of transition between galactic and extragalactic 
 cosmic rays. The exact shapes and compositions of these two components
 contains information about important parameters of powerful astrophysical
 sources and the conditions in extragalactic space as well as for the
 cosmological 
 evolution of the sources of high energy cosmic rays. Several types of 
 experimental data, including the exact shape of the ultrahigh energy
 cosmic rays, their chemical composition and their anisotropy, and 
 the fluxes of cosmogenic neutrinos have to be included in the 
 solution of this problem. 
\vspace{1pc}
\end{abstract}

\maketitle

\section{INTRODUCTION}

  In view of the expected high experimental statistics from the
 Auger Observatory~\cite{Auger} the interest in the transition
 from galactic to extragalactic cosmic rays has increased
 significantly. It is now a standard belief that above some high
 energy of order 10$^{19}$ eV all observed cosmic rays should 
 come from extragalactic sources because they cannot be contained
 in the Galaxy long enough to be accelerated~\cite{Cocconi}. For
 the rest of this paper we shall assume that the Ultra-High Energy
 Cosmic Rays (UHECR) are accelerated at powerful astrophysical objects.

  Extragalactic cosmic rays would lose energy in propagation from
 their sources to us if their sources are isotropically and homogeneously
 distributed in the Universe. The main energy loss process is the
 photoproduction interaction in the microwave background (MBR) that 
 causes the GZK effect, the steepening of the cosmic ray spectrum 
 above 6$\times$10$^{19}$ eV~\cite{GZK}. Several fits of the 
 existing experimental data have been published in recent years
 that derived different values of the most important astrophysical
 and cosmological parameters: the acceleration (injection in terms
 of cosmic ray propagation) energy spectrum of these particles and
 the maximum acceleration energy,
 their chemical composition and the cosmological evolution of their
 astrophysical sources. In the assumption that most of UHECR are
 protons injection spectra as different as E$^{-2.7}$~\cite{BGG05} and
 E$^{-2.0}$~\cite{BW03} and cosmological evolution of the type
 $(1+z)^m$ with $m$ values from 0 to 3 have been obtained.

 Other attempts~\cite{APO05a,APO05b,HST06} have assumed that extragalactic
 cosmic rays have at their sources the same mixed chemical 
 composition as the low energy galactic cosmic rays. In such a case
 the main energy loss process is the disintegration of the heavy
 nuclei mostly in interactions in MBR. Hadronic interactions
 become important only after the energy per nucleon exceeds the
 photoproduction threshold. The fits of the observed
 cosmic rays spectrum under this assumption gives an intermediate 
 $E^{-2.2-2.4}$ injection spectrum.

 In all these attempts the fits of the observations show the end
 of the galactic cosmic rays spectrum which is obtained by 
 subtraction of the propagated extragalactic spectrum from the
 experimentally observed one. This process gives some limits of
 the astrophysical parameters~\cite{ddmts05} when the subtraction
 gives unphysical negative values. 

 An important and interesting question is what such models
 predict  for the chemical composition of UHECR, which for
 our purposes we shall define as cosmic rays of energy above
 10$^{18}$~eV.
 The assumption that extragalactic cosmic rays are protons
 obviously lead to a composition that becomes very light at
 GZK and super-GZK energies. Under the assumption that 
 extragalactic cosmic rays have a mixed composition the 
 conclusions are more complicated and the question of the highest
 acceleration energy is more important.

 In terms of high energy astrophysics the location of the transition
 region can be used to estimate the strength of the cosmic ray 
 sources in different types of astrophysical objects, from 
 spiral galaxies like our own to Active Galactic Nuclei (AGN).
 In our current understanding the acceleration spectra depend
 on the strength of astrophysical shocks and the maximum
 acceleration energy is a function of the magnetic field
 strength $B$ and fluctuations $\delta B/B$. Other important 
 parameters are the cosmological evolution of the UHECR sources
 and the strength of the extragalactic magnetic fields around the
 sources and the average magnetic field in the cosmologically nearby
 Universe.

 We shall discuss several types of data and their predictions in 
 different models: UHECR spectra and chemical composition, the
 anisotropy in this energy range, and the production of 
 {\em cosmogenic}~\cite{BZ69} neutrinos by extragalactic cosmic
 rays.
 
\section{UHECR ENERGY SPECTRA}

\begin{figure}[htb]
\begin{center}
\centerline{\includegraphics[width=212pt]{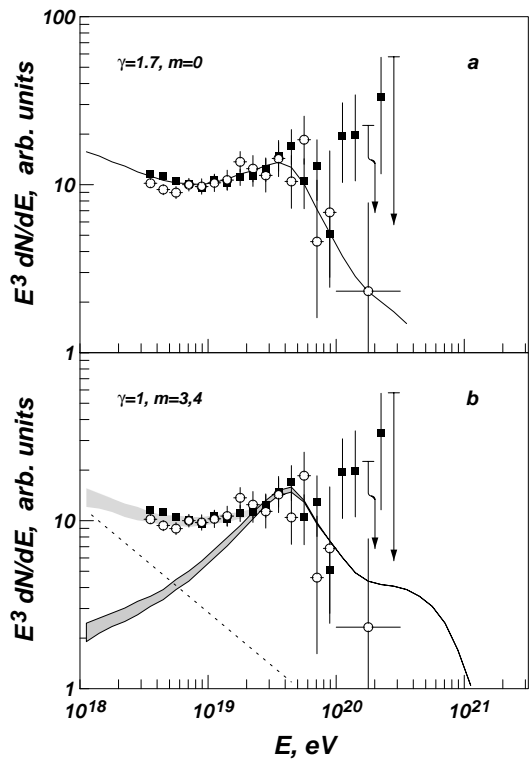}}
\vspace*{-15pt}
\caption{Comparison of two fits of the UHECR spectrum. Model {\em a}
 is from Ref.~\protect\cite{BGG05} and model {\em b} is from 
 Ref.~\protect\cite{BW03}. See text for details.
\label{ssf1a}
}
\end{center}
\vspace*{-40pt}
\end{figure}

 Figure~\ref{ssf1a} compares two different fits of the extragalactic
 cosmic ray spectra in the assumption that they are purely protons
 and that their differential acceleration spectrum is a power law
 $E^{-{\gamma+1}}$.
 The experimental data are from AGASA~\cite{AGASA} and HiRes~\cite{HiRes}
 and are normalized to each other at 10$^{19}$ eV. Since we are now
 only interested in the shape of the spectrum the exact differential
 flux at the normalization points is not important. It is obvious
 that the two experimental measurements agree well with each other
 on the shape 
 of the energy spectrum with exception of the AGASA events above
 10$^{20}$ eV. The most recent analysis of the AGASA data, presented
 at this conference~\cite{Teshima-here} decreases the energy assignment 
 of the AGASA data by 10-15\% and makes the spectrum closer to that
 of HiRes.

 Fit {\em a}~\cite{BGG05} derives an injection spectrum
 with $\gamma$ = 1.7. The dip at about 10$^{19}$ eV is due to the
 transition of
 proton energy loss to Bethe-Heitler $e^+e^-$ pair production
 to purely adiabatic loss 
 as predicted by Berezinsky\&Grigorieva~\cite{BG88}. The model 
 does not need any contribution from galactic cosmic rays to
 describe the observed cosmic ray spectrum down to 10$^{18}$ eV.
 There is also no need for cosmological evolution of the extragalactic
 cosmic ray sources, although some source evolution can be
 accommodated with a slight change of $\gamma$. The model predicts
 purely proton composition of the extragalactic cosmic rays
 and does not work as well as shown in Fig.~\ref{ssf1a} 
 if more than about 10\% of the cosmic rays at the source are 
 nuclei heavier than H.
 Fit {\em b}~\cite{BW03} does need contribution from galactic
 sources with $E^{-3.50}$ spectrum that extends well above
 10$^{19}$~eV. The extragalactic contribution is shown for two
 different cosmological evolutions with $m$ = 3 (as used in Ref.~\cite{BW03})
 and 4 (upper edge of the shaded area). The influence of the
 cosmological evolution on the cosmic ray propagation is modest
 because no protons injected at redshifts $z$ higher than 0.4 arrive
 at Earth with energy above 10$^{19}$~eV independently of their
 initial energy.

 Obviously these two models predict very differently the end of
 the galactic cosmic ray spectrum. In model {\em a} the galactic
 cosmic ray sources do not need to accelerate particles above
 10$^{18}$~eV. In model {\em b} they should be able to reach 
 energies higher by one and a half orders of magnitude. This would 
 affect very strongly the expected cosmic ray composition, as it
 will be discussed in the next Section.

 The assumption that extragalactic cosmic rays have a mixed 
 composition at acceleration gives somewhat intermediate results
 for the injection spectrum of UHECR~\cite{APO05a,APO05b,HST06}.
 The spectrum that fits the observation best has $\alpha$ values
 between 1.2 and 1.4. The chemical composition of cosmic rays 
 at Earth is also different and obviously depends on the 
 source composition.   
 
\section{CHEMICAL COMPOSITION}

  If the assumption that extragalactic cosmic rays are of mixed 
 composition similar to that of low energy galactic cosmic rays
 is correct, that would change our perception for the 
 energy dependence of the cosmic ray chemical composition.
 Before examining it let me first state that we are discussing
 the composition in terms of total energy per nucleus, not the
 composition in terms of energy per nucleon as it is usually
 presented at energies between 1 and 100 GeV/nucleon.

  Since we believe that galactic cosmic rays are accelerated 
 in stochastic processes at astrophysical shocks (as they are
 in all models except in the Dar\&DeRujula cannonball
 model~\cite{Dar,ADR}) we expect cosmic ray nuclei to have 
 maximum acceleration energy proportional to their rigidity,
 i.e. momentum/charge. Similar dependence comes out if the
 knee of the cosmic ray spectrum is due to leakage out of the
 Galaxy. The expectation is then that as soon as the maximum 
 acceleration energy for hydrogen $E_{max}^H$ is approached
 the chemical composition starts becoming heavier - the maximum
 acceleration energy for He $E_{max}^{He}$ is higher by a factor 
 of 2 (factor of 4 in the Dar\&DeRujula model). 

\begin{figure}[htb]
\begin{center}
\includegraphics[width=212pt]{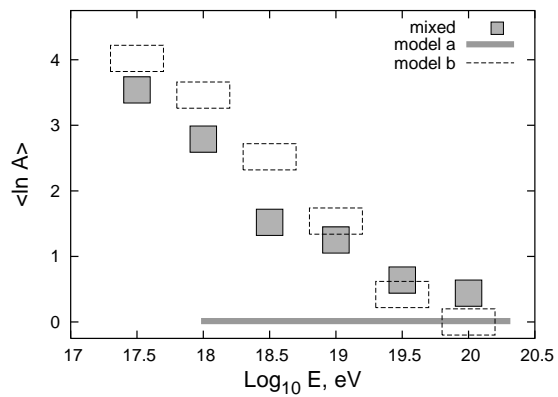}
\vspace*{-15pt}
\caption{Cosmic ray chemical composition as a function of the 
 total energy per nucleus in the three models discussed in the 
 previous section. The mixed composition model data are taken 
 Ref.~\protect\cite{APO05b}. The error bars assumed are logarithmic
 and probably lower than the true experimental ones.
\label{composi}
}
\end{center}
\end{figure}

  At the approach of $E_{max}^{Fe}$ galactic cosmic rays become
 purely iron nuclei and this is supposed to be the cosmic ray
 composition when the transition from galactic to extragalactic
 sources begins to happen. The measurements of the Kascade
 experiment~\cite{Kascade}  heavily support such behavior.
 Figure~\ref{composi} shows the
 predicted composition as a function of the total energy per 
 nucleus in the transition region in the classical cosmic ray
 units of $<ln A>$. This is very appropriate when the detection
 is through air showers with logarithmic sensitivity to both
 energy and composition.
 Model {\em a} presents the easiest case to explain. All composition 
 changes happen below 10$^{18}$~eV. We do not plot the composition below
 that energy as it is not very well defined. The results for the other
 two models are presented using our own understanding of them.
 For model {\em b} we assume that at 10$^{17.5}$~eV all cosmic rays
 are iron and assign $<ln A>$ value of 4. At higher energy we use 
 the fraction of galactic cosmic rays to iron nuclei and thus calculate
 the corresponding $<ln A>$ value. For the mixed composition we use 
 Fig.~3 of Ref.~\cite{APO05b} where the cosmic ray composition at
 Earth is plotted for $\gamma$=1.3 and $E_{max}^Z$ is given as 
 $Z\times$10$^{20.5}$~eV. 

 Surprisingly the difference in the energy dependence of $<ln A>$ 
 of model {\em b} and of the mixed composition model is not that big,
 although in the model {\em b} case the composition only consists of
 a combination of Fe and H, and in the mixed composition case we
 have five groups of nuclei. Distinguishing between these two models 
 depends on the experimental sensitivity. If the composition measurement
 uses the shower depth of maximum $X_{max}$ as a function of the
 total energy of the shower, as done in fluorescence experiments,
 the error bars shown in Fig.~\ref{composi} are approximately correct
 for $\delta X_{max}$ of about 50 g/cm$^2$. A better sensitivity
 should be able to distinguish between a two component composition
 of model {\em b} and the five component mixed composition model.

 Model {\em a} gives a very different picture, at least above
 10$^{18}$~eV, where the composition is purely Hydrogen. It is 
 definitely distinguishable from the other two models. The
 prediction that the composition does not change above 10$^{18}$~eV
 and is very light is supported by the HiRes $X_{max}$
 measurement~\cite{HiRes_comp}. What is the exact meaning of
 {\em light} composition is not known because of the differences
 between the hadronic interaction models used for data analysis.
  
 On the other hand, other experiments support a much milder
 energy dependence of the cosmic ray chemical composition, that is 
 more in line with the prediction of of models {\em b} and that
 of mixed extragalactic cosmic ray composition.

\section{ANISOTROPY}

 Low energy cosmic rays diffuse in the magnetic fields of the Galaxy
 and lose memory of the location of their sources. The anisotropies
 are very small, well below 1\% and are difficult to measure.
 The measurement of a small anisotropy with air shower experiments
 requires an exact knowledge of the experimental acceptance and of
 the lifetime of the shower array. 

 At higher energy things are supposed to change as the cosmic ray
 rigidity increases and their diffusion in the Galaxy becomes faster.
 The first question we still cannot answer is of the rigidity
 (energy) dependence of the cosmic ray diffusion coefficient.
 From the ratio of secondary to primary cosmic rays an energy
 dependence of $E^{-0.5-0.6}$ is derived. If this dependence is extended
 by seven orders of magnitude then galactic UHE protons should
 show very strong anisotropy that is not observed. The two spots
 with increased cosmic ray flux identified by the AGASA group~\cite{A_aniso}
 (one in the general direction of the Galactic center and one in the
 direction of Orion) are not confirmed by other experiments.
 Theoretically we expect Kolmogorov turbulence in the Galaxy, that
 should under some circumstances give $E^{1/3}$ energy dependence.

 One way to explain the low anisotropy in the beginning of the 
 transition region is to assume that all galactic cosmic rays are
 heavy nuclei. If they were indeed heavy nuclei
 the average particle rigidity would 
 significantly decrease and would maintain the high isotropy
 of the galactic cosmic rays. In the transition region, however,
 the cosmic ray chemical composition becomes lighter and 
 correspondingly we expect to see some degree of anisotropy.
\begin{figure}[htb]
\begin{center}
\vspace*{-20pt}
\includegraphics[width=212pt]{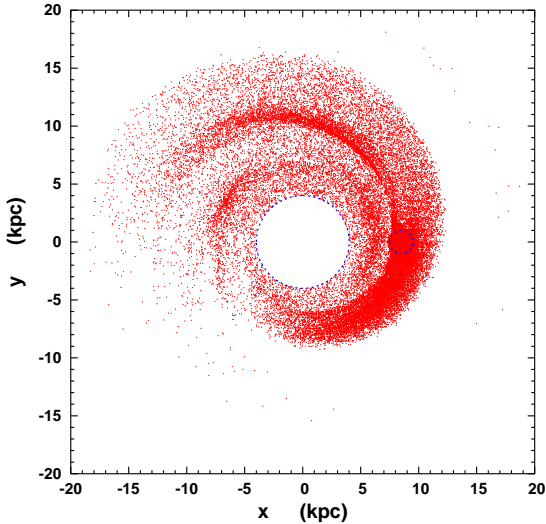}
\vspace*{-15pt}
\caption{Location of sources from which isotropically emitted protons
 in the galactic plane reach the Earth. See text for the details of 
 the calculation.
\label{sources}
}
\end{center}
\vspace*{-20pt}
\end{figure}
 Figure~\ref{sources} shows the location of the sources of cosmic ray
 protons that reach the Earth in a propagation calculation.
 The source distribution is assumed
 to be $R_{GC}^{-1}$ where $R_{GC}$ is the galactocentric distance.
 There are no sources inside the 4 kpc circle. Cosmic rays are
 isotropically  injected in the galactic plane on a $E^{-2.7}$ spectrum
 with energies between 4$\times$10$^{17}$ and 2$\times$10$^{18}$~eV.
 The magnetic field model is BSS with random field and a magnetic
 dipole in the center of the Galaxy. The Earth is approximated by a sphere 
 of radius 1 kpc to speed the calculation.

 The figure shows the general behavior of protons in this energy range.
 They do not diffuse, rather follow the magnetic field lines. Because
 of the magnetic turbulence the protons occasionally jump from one
 magnetic field line to another and may change the direction of their
 propagation depending on the pitch angle. 

 When these protons arrive at Earth their arrival direction does not
 point at their source, they have rather a distribution that peaks
 in the direction of the local field line. Since we expect higher
 source density inside the solar circle, the expected anisotropy
 may peak in the general direction of Orion. Exercises like that one
 do create relatively strong anisotropy, much higher than observed.

 One may complain that in Fig.~\ref{sources} we are looking at
 particles that are accelerated in the galactic plane while in the
 case of extragalactic protons we will be dealing with particles
 that isotropically enter into the Galaxy and may not create any
 anisotropy. The truth is that calculations like the one that
 generated Fig.~\ref{sources} show that protons of energy above
 10$^{16}$ eV emitted in the galactic plane leave the Galaxy much
 easier in the direction of the galactic poles that after propagation
 parallel to the galactic plane. In the entrance of the Galaxy 
 we would expect the same effect - more extragalactic protons should
 arrive from absolute high galactic latitude $|b|$ than from low one.

 This would not be the classical anisotropy that surveys the 
 arrival direction as a function of the galactic longitude, but
 it should still be significant effect. The details of the effect
 depend strongly on the galactic magnetic field model~\cite{JAMS}
 and on details of the calculation such as the dimension of the
 `Earth' in the calculation and of the propagation step size.

 I am convinced that in the case of high experimental statistics
 the anisotropy in the transition region between galactic and
 extragalactic cosmic rays deserves a careful study. A part of
 it is theoretical. We should propagate particles of different
 rigidity in the Galaxy and attempt to understand their general
 behavior. Analytic solutions of the diffusion equation are not
 any more suitable for this problem. Particle trajectory has to be 
 numerically solved, most likely in a Monte Carlo fashion and in
 detailed enough magnetic field models. Such models should be
 tested to match analytic calculations when applied to the appropriate
 simplified models.
 
\section{COSMOGENIC NEUTRINOS}

 Cosmogenic neutrinos are produced in the same photoproduction
 interactions of the UHECR protons that create the GZK effect.
 They were first proposed by Berezinsky\&Zatsepin in 1969~\cite{BZ69}
 and were the subject of many calculations afterward.
 Cosmogenic neutrinos are often considered a `guaranteed source'
 of ultrahigh energy neutrinos. They are indeed guaranteed, since
 we know UHECR exist, but their flux is unknown.

 An essential quality of neutrinos is that they have a low 
 interaction cross section. This makes neutrino detection a
 difficult problem that requires huge detectors of at least
 km$^3$ scale. Such detectors are now in the stage of
 planning~\cite{KM3Net} and construction~\cite{IceCube,ANITA}.
 The question of the cosmogenic neutrino flux and of its 
 relation to other data from ultra-high astrophysics experiment
 is very timely.

 The low interaction cross section is not only a deficiency.
 While protons emitted at redshifts $z$ exceeding 0.4 do not
 reach us with energy above 10$^{19}$~eV the cosmogenic 
 neutrino production peaks at redshifts exceeding 2 for 
 $(1+z)^3$ cosmological evolution of the UHECR sources.
 This is the main link between the extragalactic UHECR
 and cosmogenic neutrinos. The detection of cosmogenic 
 neutrinos may help the degeneracy in modeling of the
 extragalactic cosmic rays spectra shown in Fig.~\ref{ssf1a}.

\begin{figure}[htb]
\begin{center}
\includegraphics[width=212pt]{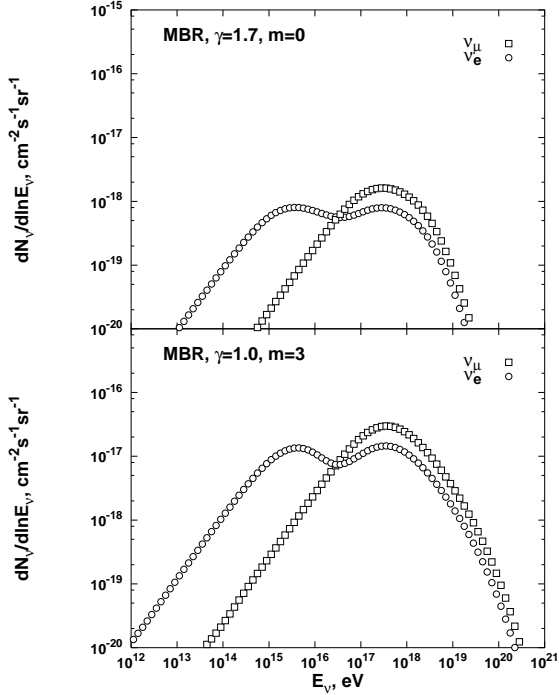}
\vspace*{-15pt}
\caption{Fluxes of cosmogenic neutrinos generated in proton interactions
 in the microwave background~\protect\cite{ESS01} in the two models
 of the UHECR spectrum. Squares show the sum of \protect$\nu_\mu +
 \bar{\nu}_\mu$ and the circles - the sum of \protect$\nu_e + \bar{\nu}_e$.
\label{mbr_nu}
}
\end{center}
\vspace*{-20pt}
\end{figure}

 The point is that models that use flat cosmic ray injection
 spectrum and require strong cosmological evolution of the
 UHECR sources, such as model {\em b} would produce significantly
 more cosmogenic neutrinos than steep injection spectrum models
 with no cosmological evolution. Such a comparison between the
 two models is shown in Figure~\ref{mbr_nu}.
 The figure shows cosmogenic neutrino fluxes generated by proton 
 interactions in MBR with the injection spectrum and cosmological
 evolution of the two models as indicated in the two panels.
 The injection spectra of the two models are normalized to 
 the UHECR differential flux at 10$^{19}$ eV.
 Before discussing the magnitude of the two fluxes I want to 
 attract your attention to the flux of electron neutrinos and
 antineutrinos that exhibits two maxima separated by about two orders 
 of magnitude of energy. The higher energy one is due to the
 $\pi \rightarrow \nu_\mu + \bar{\nu}_\mu + \nu_e$ decay final
 state and consists mostly of electron neutrinos. It peaks at
 the same energy where $\mu_\mu$ and $\bar{\nu}_\mu$ do. The lower 
 energy maximum is due to electron antineutrinos from neutron
 decay. Comparison between the neutron interaction length in MBR
 and their decay length show that all neutrons of energy below 
 10$^{20.5}$~eV decay rather than interact.

 Model {\em b} generates much higher cosmogenic neutrino fluxes
 than model {\em a} because of two reasons that contribute
 roughly the same increase of the cosmological neutrinos.
 Firstly, it uses much flatter injection spectrum $E^{-2.0}$
 which means equal amount of energy per decade. It thus
 contains many more particles above the photoproduction
 interaction threshold with in the MBR is about
 3$\times$10${19}$~eV and, of course, decreases as $(1+z)^{-1}$.

 The other reason is that model {\em b} employs a strong
 cosmological evolution of the cosmic ray sources. This
 increases by $\sqrt{3}$ the number of particles injected
 at redshift of 2, but in addition increases by a factor
 of three the number of particles above the interaction threshold.
 This way the total neutron flux is increased by the cosmological
 evolution of the sources by a factor of five. 

 The difference in the peak values of the cosmological neutrinos
 generated by the two models is more than one order of magnitude.
 In practical terms this means that model {\em b} generates 
 fluxes that are in principle detectable by IceCube at the
 rate of roughly less than one event per year, while model {\em a} 
 generates undetectable fluxes of cosmogenic neutrinos in km$^3$
 detectors.

  Ice and water neutrino detectors are generally not
 very suitable for cosmogenic neutrino detection. Much better 
 strategies for these UHE neutrinos are the radio and acoustic
 detectors that have very high detection threshold, but also
 will have higher effective volume. The other option are
 giant air shower arrays such as Auger, that can reach 
 effective volume of 30 km$^3$ and, with sufficiently low
 threshold (10$^{18}$~eV) could detect several events per 
 year. 

 Cosmogenic neutrinos are also generated in the mixed composition
 scenario~\cite{HTS05,ABO05}. Since the major energy loss
 process is the disintegration of heavy nuclei, the main flux 
 component of $\bar{\nu}_e$ from neutron decay. The 
 $\bar{\nu}_e$ flux exceeds by a factor of 5 the sum of the 
 neutrino fluxes of all other flavors in Ref.~\cite{ABO05}.
 The absolute magnitude depends again mainly on the injection spectrum
 and the cosmological evolution of the sources.
  
\subsection{Production in the infrared/optical background}

 The difference between different models  is somewhat decreased 
 when photon fields different from MBR are considered. 
 Several calculations that include the infrared/optical background
 (IRB) have been performed.  The number density of IRB is of order
 1 and varying by about a factor of 2 in different models, but 
 its energy spectrum extends to energies as high as and exceeding
 1 eV. What that means is that extragalactic particles of much
 lower energy would interact in IRB and will generate neutrinos.
 Reference~\cite{SDMMS06} shows the neutrino yields generated by
 protons of energy as low as 10$^{18}$~eV. This yield is small but
 has to be weighted by the much higher number of particles at that
 energy - almost 1,000 higher that that of 3$\times$10$^{19}$~eV 
 for a flat $E^{-2}$ injection spectrum. A more recent calculation
 that also includes UV photons~\cite{AAB06} employs 
 interactions of 10$^{17}$ eV nucleons that have to be weighted
 still higher.
\begin{figure}[htb]
\begin{center}
\includegraphics[width=212pt]{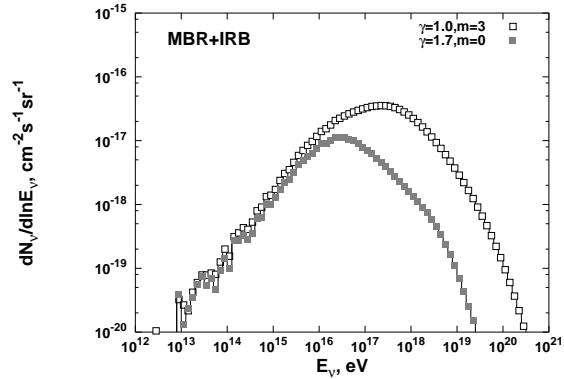}
\vspace*{-15pt}
\caption{Fluxes of cosmogenic neutrinos generated in
 the MBR and IRB. Only the sum of \protect$\nu_\mu+\bar{\nu}_\mu$
 are shown.
\label{all_nu}
}
\end{center}
\vspace*{-20pt}
\end{figure}
 Figure~\ref{all_nu} shows the $\nu_\mu + \bar{\nu}_\mu$ cosmogenic
 neutrino fluxes  from interactions in the MBR and IRB~\cite{SDMMS06}
 by the two models. The difference in the peak values is now slightly
 less than a factor of three. The steep injection spectrum model
 {\em a} has significantly more protons that interact in the IRB and
 for that reason partially compensates for the lack of cosmological
 evolution of the sources. On the other hand, since the average
 neutrino energy is substantially lower and the neutrino interaction 
 cross section increases with energy, the event rates produced
 by the two models would not change much. 
 If one uses steep injection spectrum combined with an appropriate
 cosmological evolution of the sources that inclusion of the IRB 
 target makes a big difference.
 
\section{DISCUSSION}

 The transition region between galactic and extragalactic cosmic 
 rays is currently not very well defined and studied. It has
 a significant astrophysical importance, because the energy content
 of the extragalactic cosmic rays, as well as their injection
 spectrum and composition, will define much better the type of 
 their sources. In addition, a derivation of the cosmological
 evolution of the sources will not only restrict the number of
 available scenarios, but also provide an additional measure
 of the cosmological evolution of powerful astrophysical objects.

 This will not be easy to do even with the expected much higher
 statistics and detection precision of the Auger Observatory.
 We should employ all available types of measurements to solve
 this problem as well as it is necessary. These measurements 
 include the cosmic ray composition in the transition region 
 and above it, the cosmic ray anisotropies, and the possible
 detection of cosmogenic neutrinos.

 The first experimental result on the shape of this transition
 presented by the Fly's Eye group~\cite{Birdetal93} used the
 simultaneous change of the cosmic ray composition and energy
 spectrum shape. We have to do the same now - to combine 
 all observational data, look for consistency between different
 sets, and with models of the extragalactic cosmic rays origin.

 The possible detection of cosmogenic neutrinos would be a
 powerful test if we succeed in collecting a reasonable statistics
 of such events. It is unlikely this will happen very soon, but
 the expanding efforts for designing and building detectors for
 ultrahigh energy neutrinos are very encouraging.

 {\bf Acknowledgments} This talk is based on work performed in 
 collaboration with D.~Seckel, D.~DeMarco, J.~Alvarez-Mu\~{n}iz,
 R.~Engel and others. Partial support of my work comes from 
 US DOE Contract DE-FG02 91ER 40626 and NASA Grant ATP03-0000-0080.


\begin{thebibliography}{99}
\bibitem{Auger} see {\em http://auger.org}
\bibitem{Cocconi} G.~Cocconi, Nuovo Cim., 3:1433 (1956)
\bibitem{GZK} K.~Greisen, Phys. Rev. Lett., 10:146 (1966);
 G.T.~Zatsepin \& V.A.~Kuzmin, Pisma Zh. Exp. Theor. Phys. 7:181 (1966
\bibitem{BGG05} V.~Berezinsky, A.Z.~Gazizov \& S.I.~Grigorieva,
 Phys. Lett. B612:147 (2005); The idea was first presented by
 the same authors in astro-ph/0204357 and developed in several
 other papers, most recently in R.~Aloisio, V.~Berezinsky, P.~Blasi et al.,
 astro-ph/0608219
\bibitem{BW03} J.N.~Bahcall \& E.~Waxman, Phys. Lett. B556:1 (2003)
\bibitem{APO05a} D.~Allard, E.~Parizot, A.~Olinto et al., 
 A\&A, 443:L29 (2005)
\bibitem{APO05b} D.~Allard, E.~Parizot \& A.~Olinto, {\em astro-ph/0512345} 
\bibitem{HST06} D.~Hooper, S.~Sarkar \& A.M.~Taylor, {\em astro-ph/0608085}
\bibitem{ddmts05} D.~DeMarco \& T.~Stanev, Phys. Rev. D72:081301 (2005)
\bibitem{BZ69} V.S.~Berezinsky \& G.T.~Zatsepin, Phys. Lett., 28B:423 (1969)
\bibitem{AGASA} M.~Takeda et al., Astropart. Phys., 19:447 (2003)
\bibitem{HiRes} R.U.~Abbasi et al. (HiRes Collaboration),
 Phys.Rev.Lett.92:151101 (2004)
\bibitem{Teshima-here} M.~Teshima, talk at this meeting.
\bibitem{BG88} V.S.~Berezinsky \& S.I.~Grigorieva, Astron. Astrophys.,
 199:1 (1988) 
\bibitem{Dar} see talk at this meeting, also A.~Dar \& A.~DeRujula,
 {\em hep-ph/0606199} 
\bibitem{ADR} see talk at this meeting, also {\em hep-ph/0606199}
\bibitem{Kascade} T.~Antoni et al. (Kascade collaboration), Astropart. Phys.,
 24:1 (2005)
\bibitem{HiRes_comp} R.U. Abbasi et al. (HiRes Collaboration).
  Ap.~J.622:910-926 (2005)
\bibitem{A_aniso} N.~Hayashida et al. (AGASA Collaboration),
 Astropart. Phys., 10:303 (1999)
\bibitem{JAMS} J.~Alvarez-Mu\~{n}iz \& T.~Stanev, in Proc Aspen Workshop
 on the End of the Galactic Cosmic Ray Spectrum, {\em astro-ph/0507273} 
\bibitem{KM3Net} {\em http://km3net.org}
\bibitem{IceCube} {\em http://icecube.wisc.edu}
\bibitem{ANITA} S.W.~Barwick et al. (ANITA Collaboration),
 Phys.Rev.Lett.96:171101 (2006)
\bibitem{ESS01} R.~Engel, D.~Seckel \& T.~Stanev, Phys. Rev. D64:093010 (2001)
\bibitem{HTS05} D.~Hooper, A.~Taylor \& S.~Sarkar, Astropart. Phys.,
 23:11 (2005)]
\bibitem{ABO05} M.~Ave, N.~Busca, A.~Olinto et al., Astropart. Phys.,
 23:19 (2005)]
\bibitem{SDMMS06} T.~Stanev, D.~DeMarco, M.~Malkan \& F.~Stecker,
 Phys. Rev. D73:043003 (2006)
\bibitem{AAB06} D.~Allard, M.~Ave, N.~Busca et al., {\em astro-ph/0605327}
\bibitem{Birdetal93} D.J.~Bird et al. (HiRes Collaboration),
 Phys. Rev. Lett. 71:3401 (1993) 
\end{thebibliography}
\end{document}